\documentclass[aps,english,showpacs,twocolumn]{revtex4-1}
\usepackage{amssymb}
\usepackage{amsmath}
\usepackage{graphicx}
\usepackage{epsfig}

\begin{document}

\title{Hermitian scattering behavior for the non-Hermitian scattering center}
\author{L. Jin and Z. Song}
\email{songtc@nankai.edu.cn}
\affiliation{School of Physics, Nankai University, Tianjin 300071, China}

\begin{abstract}
{We study the scattering problem for the non-Hermitian scattering center,
which consists of two Hermitian clusters with anti-Hermitian couplings
between them. Counterintuitively, it is shown that it acts as a Hermitian
scattering center, satisfying $\left\vert r\right\vert ^{2}+\left\vert
t\right\vert ^{2}=1,$ i.e., the Dirac probability current is conserved, when
one of two clusters is embedded in the waveguides. This conclusion can be
applied to an arbitrary parity-symmetric real Hermitian graph with
additional $\mathcal{PT}$-symmetric potentials, which is more feasible in
experiment. Exactly solvable model is presented to illustrate the theory.
Bethe ansatz solution indicates that the transmission spectrum of such a
cluster displays peculiar feature arising from the non-Hermiticity of the
scattering center.}
\end{abstract}

\maketitle


\section{Introduction}

\label{sec_intro} A non-Hermitian Hamiltonian is usually endowed with the
physical meaning when it possesses entirely real quantum mechanical energy
spectrum and the complex extension of the conventional quantum mechanics, a
parity-time ($\mathcal{PT}$) symmetric quantum theory, has been well
developed \cite{Bender 98,Bender 99,Dorey 01,Bender 02,A.M43,A.M,A.M36,Jones}
since the seminal discovery by Bender \cite{Bender 98}. Such a theory gives
the pseudo-Hermitian Hamiltonian a physical meaning via its corresponding
Hermitian counterparts \cite{A.M38,A.M391,A.M392}, which has an identical
spectrum. The metric-operator theory outlined in Ref. \cite{A.M} provides a
mapping of such a pseudo-Hermitian Hamiltonian to an equivalent Hermitian
Hamiltonian. Thus, most of the studies focused on the quasi-Hermitian
system, or unbroken $\mathcal{PT}$-symmetric region \cite%
{Joglekar82,Joglekar83}. However, the obtained equivalent Hermitian
Hamiltonian is usually quite complicated \cite{A.M,JLPT}, involving
long-range or nonlocal interactions, which is hardly realized in practice.

Experimentally, the $\mathcal{PT}$\ symmetry is of great relevance to the
technological applications based on the fact that the imaginary potential
could be realized by complex index in optics \cite%
{Bendix,Keya,YDChong,LonghiLaser}.\ Furthermore, the $\mathcal{PT}$ optical
potentials can be realized through a judicious inclusion of index guiding
and gain/loss regions. Such non-Hermitian systems are not isolated but
usually embedded in the large Hermitian waveguides. Pure imaginary potential
as a scattering center breaks the conservation of the flow of probability
\cite{JLCTP}. Thus, it is interesting to investigate what happens when the
non-Hermitian system is with balanced gain and loss as a scattering
center, and much effort devoted to such a topic is based on the framework of
$\eta $-metric \cite{ZAhmed,MZnojil,HJones,Kivshar}.

In this paper, we study the scattering problem for the non-Hermitian
scattering center based on the configurations involving two arbitrary
Hermitian networks coupled with anti-Hermitian interaction. It is shown that
for any scattering state of such a non-Hermitian system, the Dirac
probability current is always conserved at any degree of the
non-Hermiticity. We apply such a rigorous result to the system with $%
\mathcal{PT}$-symmetric potentials, which is more feasible in experiment.

This paper is organized as follows. Section \ref{sec_formalism} presents the
exact analytical solution of the scattering problem for the concerned
non-Hermitian scattering center. Section \ref{sec_PT_Poten} is the
application of the rigorous result to the system with $\mathcal{PT}$%
-symmetric potentials. Section \ref{sec_illus} consists of an exactly
solvable example to illustrate our main idea. Section \ref{sec_summary} is
the summary and discussion.

\section{Model and formalism}

\label{sec_formalism} In general, a non-Hermitian Hamiltonian $H$\ is
related by a similarity transformation to an equivalent Hermitian
Hamiltonian $h$. Such a connection is valid within the so called unbroken
symmetric region. However, when a non-Hermitian system interacts with other
Hermitian system, such a region loses its physical meaning: On the one hand,
the unbroken symmetric region is shifted in the whole non-Hermitian system.
On the other hand, it may act as a Hermitian system in the scattering
problem without the restriction on the\ degree of the non-Hermiticity. In
this section, we will investigate the latter situation.


\begin{figure}[tbp]
\includegraphics[ bb=64 462 529 686, width=8.5 cm, clip]{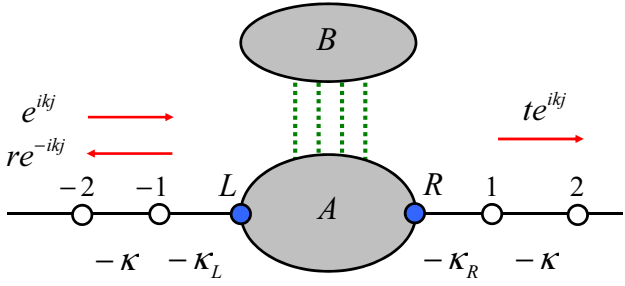}
\caption{((Color online) Schematic illustration of the configuration of
the concerned network. It consists of two arbitrary graphs of the Hermitian
tight-binding networks $A$ and $B$ (shadow) with one of them connecting to
two semi-infinite chains as the waveguides at the joint sites $L$ and $R$.
The non-Hermiticity of the whole scattering center $A+B$\ arises from the
anti-Hermitian interaction (dished lines) between them. It is shown that the
non-Hermitian scattering center acts as a Hermitian\ one, preserving the
Dirac probability current.} \label{fig1}
\end{figure}

The Hamiltonian of the concerned scattering tight-binding network has the
form

\begin{equation}
H=H_{L}+H_{R}+H_{C},  \label{H}
\end{equation}%
where%
\begin{eqnarray}
H_{L} &=&-\kappa \sum_{j=-1}^{-\infty }\left\vert j\right\rangle
_{L}\left\langle j-1\right\vert -g_{L}\left\vert L\right\rangle
_{L}\left\langle -1\right\vert +\text{H.c.},  \label{H_L} \\
H_{R} &=&-\kappa \sum_{j=1}^{+\infty }\left\vert j\right\rangle
_{R}\left\langle j+1\right\vert -g_{R}\left\vert R\right\rangle
_{R}\left\langle 1\right\vert +\text{H.c.},  \label{H_R}
\end{eqnarray}%
represent the left ($H_{L}$) and right ($H_{R}$) waveguides with real $%
\kappa $\ and%
\begin{equation}
H_{C}=H_{A}+H_{B}+H_{AB}+H_{BA},  \label{H_c}
\end{equation}%
describes a non-Hermitian network as a scattering center. Here $\left\vert
L\right\rangle $ and $\left\vert R\right\rangle $ denote the sites state on
the joint sites on the network $A$, which are simply taken as $\left\vert
L\right\rangle =\left\vert 1\right\rangle _{A}$ and $\left\vert
R\right\rangle =\left\vert N_{A}\right\rangle _{A}$ without losing the
generality. The subgraphs%
\begin{eqnarray}
H_{A} &=&\sum_{i,j=1}^{N_{A}}\left( H_{A}\right) _{ij}\left\vert
i\right\rangle _{A}\left\langle j\right\vert ,  \label{H_A} \\
H_{B} &=&\sum_{i,j=1}^{N_{B}}\left( H_{B}\right) _{ij}\left\vert
i\right\rangle _{B}\left\langle j\right\vert ,  \label{H_B}
\end{eqnarray}%
are arbitrary Hermitian networks, i.e., $H_{A}^{\dag }=H_{A}$, and $%
H_{B}^{\dag }=H_{B}$, while the coupling between them is anti-Hermitian,
i.e., $H_{AB}^{\dagger }=-H_{BA}$.%
\begin{equation}
H_{AB}=\sum_{i=1}^{N_{A}}\sum_{j=1}^{N_{B}}\left( H_{AB}\right)
_{ij}\left\vert i\right\rangle _{AB}\left\langle j\right\vert ,  \label{H_AB}
\end{equation}%
then the scattering center\ with respect to the basis $\left\{ \left\vert
i\right\rangle _{A},\left\vert i\right\rangle _{B}\right\} $ is in the form
of%
\begin{equation}
H_{C}=\left(
\begin{array}{cc}
H_{A} & H_{AB} \\
-H_{AB}^{\dag } & H_{B}%
\end{array}%
\right) .  \label{H_C_Matrix_form}
\end{equation}%
The non-Hermiticity of $H_{C}$\ arises from this anti-Hermitian term. The
non-Hermitian Hamiltonian $H_{C}$\ may have fully real spectrum or not. In
the following, we will show that it always acts as a Hermitian scattering
center no matter the reality of the spectrum.

For an incident plane wave with momentum $k$ incoming from the left
waveguide $L$ with energy $E=-2\kappa \cos k$, the scattering wave function $%
\left\vert \psi _{k}\right\rangle $ can be obtained by the Bethe ansatz
method. The wave function has the form

\begin{equation}
\left\vert \psi _{k}\right\rangle =\sum_{j=-1}^{-\infty }f_{j}\left\vert
j\right\rangle _{L}+\sum_{j=1}^{N_{A}}\alpha _{j}\left\vert j\right\rangle
_{A}+\sum_{j=1}^{N_{B}}\beta _{j}\left\vert j\right\rangle
_{B}+\sum_{j=1}^{+\infty }f_{j}\left\vert j\right\rangle _{R},  \label{Psi_k}
\end{equation}%
where the scattering wavefunction $f_{j}$ is in form of

\begin{equation}
f_{j}=\left\{
\begin{array}{c}
e^{ikj}+re^{-ikj},\left( j\leqslant -1\right) \\
te^{ikj},\left( j\geqslant 1\right)%
\end{array}%
\right. .  \label{f_i}
\end{equation}%
Here $r$, $t$ are the reflection and transmission coefficients of the
incident wave, which is what we concern only in this paper. Substituting the
wavefunction $\left\vert \psi _{k}\right\rangle $ into the Schr\"{o}dinger
equation

\begin{equation}
H\left\vert \psi _{k}\right\rangle =E\left\vert \psi _{k}\right\rangle ,
\label{S-eq}
\end{equation}%
the explicit form of the Schr\"{o}dinger equations in the truncated Hilbert
space spanned by the basis $\left\{ \left\vert j,j\in \left[ 1,N_{A}\right]
\right\rangle _{A},\left\vert j,j\in \left[ 1,N_{B}\right] \right\rangle
_{B}\right\} $ can be expressed in the following matrix equation form%
\begin{equation}
\Delta \left(
\begin{array}{c}
\alpha _{1} \\
\vdots \\
\alpha _{j} \\
\vdots \\
\alpha _{N_{A}} \\
\beta _{1} \\
\vdots \\
\beta _{j} \\
\vdots \\
\beta _{N_{B}}%
\end{array}%
\right) =\left(
\begin{array}{c}
g_{L}f_{-1} \\
0 \\
\vdots \\
0 \\
g_{R}f_{1} \\
0 \\
\vdots \\
0 \\
\vdots \\
0%
\end{array}%
\right) ,  \label{truncate}
\end{equation}%
where $\Delta $ is an $\left( N_{A}+N_{B}\right) \times $ $\left(
N_{A}+N_{B}\right) $ matrix defined by

\begin{equation}
\Delta =\left(
\begin{array}{cc}
H_{A}-E & H_{AB} \\
-H_{AB}^{\dag } & H_{B}-E%
\end{array}%
\right) ,  \label{Delta}
\end{equation}%
From the reduced Schr\"{o}dinger equation of Eq. (\ref{truncate}), we obtain

\begin{eqnarray}
\alpha _{1} &=&\left( \Delta ^{-1}\right) _{11}g_{L}f_{-1}+\left( \Delta
^{-1}\right) _{1N_{A}}g_{R}f_{1},  \label{S_eq_1} \\
\alpha _{N_{A}} &=&\left( \Delta ^{-1}\right) _{N_{A}1}g_{L}f_{-1}+\left(
\Delta ^{-1}\right) _{N_{A}N_{A}}g_{R}f_{1},  \notag
\end{eqnarray}%
Here $\Delta ^{-1}$ is the inverse of matrix $\Delta $, with the element
being expressed as

\begin{equation}
\left( \Delta ^{-1}\right) _{ij}=\frac{C_{ji}}{\det \left( \Delta \right) }=%
\frac{\left( -1\right) ^{i+j}\det \left( M_{ji}\right) }{\det \left( \Delta
\right) },  \label{minor}
\end{equation}%
in term of the matrix of cofactors $C_{ij}$. Here $M_{ij}$\ is the matrix
obtained by deleting the $i$th row and $j$th column from the matrix $\Delta $%
. On the other hand, the Schr\"{o}dinger equations for the sites of the
waveguides\ connecting to the joints of the scattering center are%
\begin{eqnarray}
-\kappa f_{-2}-g_{L}^{\ast }\alpha _{1} &=&Ef_{-1},  \label{S_eq_2} \\
-\kappa f_{2}-g_{R}^{\ast }\alpha _{N_{A}} &=&Ef_{1},  \notag
\end{eqnarray}%
which lead to

\begin{equation}
\alpha _{1}=\frac{\kappa }{g_{L}^{\ast }}\left( 1+r\right) \text{, }\alpha
_{N_{A}}=\frac{\kappa }{g_{R}^{\ast }}t.
\end{equation}%
Then associating with Eqs. (\ref{S_eq_1}), we have

\begin{eqnarray}
r &=&(-b\widetilde{b}+ac-ae^{-ik}-ce^{ik}+1)/\eta ,  \label{r} \\
t &=&i2\widetilde{b}\sin k/\eta ,  \label{t}
\end{eqnarray}%
where

\begin{eqnarray}
\eta &=&(b\widetilde{b}-ac)e^{i2k}+\left( a+c\right) e^{ik}-1,  \notag \\
a &=&\left( \Delta ^{-1}\right) _{11}\left\vert g_{L}\right\vert ^{2}/\kappa
\text{, }c=\left( \Delta ^{-1}\right) _{N_{A}N_{A}}\left\vert
g_{R}\right\vert ^{2}/\kappa ,  \label{abc} \\
b &=&\left( \Delta ^{-1}\right) _{1N_{A}}g_{L}^{\ast }g_{R}/\kappa \text{, }%
\widetilde{b}=\left( \Delta ^{-1}\right) _{N_{A}1}g_{L}g_{R}^{\ast }/\kappa .
\notag
\end{eqnarray}%
One can determine the unknown coefficients $a$, $b$, $\widetilde{b}$, $c$\
and $\eta $\ through the matrix $\Delta $\ by requiring that invertible
matrix $\left( H_{A}-E\right) $\ or $\left( H_{B}-E\right) $\ exists.

In the Appendix, we will show that
\begin{equation}
\left( \Delta ^{-1}\right) _{ij}=\left( \Delta ^{-1}\right) _{ji}^{\ast }
\label{Delta_ij}
\end{equation}%
for $i,j\in \lbrack 1,N_{A}]$, or more explicitly for special cases
\begin{subequations}
\label{inverse_delta_rela}
\begin{eqnarray}
\left( \Delta ^{-1}\right) _{11} &=&\left( \Delta ^{-1}\right) _{11}^{\ast },
\label{real Delta_a} \\
\left( \Delta ^{-1}\right) _{N_{A}N_{A}} &=&\left( \Delta ^{-1}\right)
_{N_{A}N_{A}}^{\ast },  \label{real Delta_b} \\
\left( \Delta ^{-1}\right) _{1N_{A}} &=&\left( \Delta ^{-1}\right)
_{N_{A}1}^{\ast },  \label{conjugate}
\end{eqnarray}%
which indicate that both $a$ and $c$ are real, and $\widetilde{b}=b^{\ast }$%
. It is somewhat surprising that we get the conclusion from Eqs. (\ref{r}), (%
\ref{t}), (\ref{abc}) and (\ref{inverse_delta_rela}) that

\end{subequations}
\begin{equation}
\left\vert r\right\vert ^{2}+\left\vert t\right\vert ^{2}=1,  \label{t2+r2=1}
\end{equation}%
which is common phenomenon\ in a Hermitian system but surprising in a
non-Hermitian system.

\section{$\mathcal{PT}$-symmetric potentials}

\label{sec_PT_Poten} The accessible setup of non-Hermitian system in the lab
is the $\mathcal{PT}$-symmetric potentials, which can be realized through a
judicious inclusion of index guiding and gain/loss regions. In the
following, we will apply the obtained result to the system with the $%
\mathcal{PT}$-symmetric potentials, in which the $\mathcal{PT}$-symmetrical
axis is along the waveguides.

The geometry of the scattering center contains $N_{1}+2N_{2}$ sites and
possesses the following symmetry,%
\begin{equation}
\begin{array}{l}
\mathcal{P}:\left\vert j\right\rangle _{c}\longrightarrow \left\vert
j\right\rangle _{c},\text{ }\left( j\in \left[ 1,N_{1}\right] \right) \\
\mathcal{P}:\left\vert j\right\rangle _{c}\longrightarrow \overline{%
\left\vert j\right\rangle }_{c}=\left\vert N_{2}+j\right\rangle _{c},\text{ }%
\left( j-N_{1}\in \left[ 1,N_{2}\right] \right)%
\end{array}
\label{P}
\end{equation}%
with the joint points $L,$ $R$ $\in \left[ 1,N_{1}\right] $, where $%
\overline{\left\vert j\right\rangle }_{c}$ is the mirror symmetric
counterpart of state $\left\vert j\right\rangle _{c}$. We define\ the
Hamiltonian of the center has the form

\begin{eqnarray}
&&H_{\mathcal{PT}}=\sum_{i,j=1,\left( i<j\right) }^{N_{1}+2N_{2}}\kappa
_{ij}\left\vert i\right\rangle _{c}\left\langle j\right\vert +\text{H.c.}
\label{H_PT} \\
&&+\sum_{j=1}^{N_{1}}U_{j}\left\vert j\right\rangle _{c}\left\langle
j\right\vert +\sum_{j=N_{1}+1}^{N_{1}+N_{2}}(V_{j}\left\vert j\right\rangle
_{c}\left\langle j\right\vert +V_{j}^{\ast }\overline{\left\vert
j\right\rangle }_{c}\overline{\left\langle j\right\vert }),  \notag
\end{eqnarray}%
where\ $\kappa _{ij}$ and $U_{j}$ are real. In the Hilbert space spanned by
basis $\left\{ \left\vert j\right\rangle _{c}\right\} $ $\left( j\in \left[
1,N_{1}+2N_{2}\right] \right) $, the matrix of the Hamiltonian $H_{\mathcal{%
PT}}$ has the form

\begin{equation}
H_{\mathcal{PT}}=\left(
\begin{array}{ccc}
H_{\gamma } & H_{\gamma \alpha } & H_{\gamma \alpha } \\
H_{\gamma \alpha }^{\dagger } & H_{\alpha }+H_{\delta } & H_{\alpha \beta }
\\
H_{\gamma \alpha }^{\dagger } & H_{\alpha \beta } & H_{\alpha }-H_{\delta }%
\end{array}%
\right) ,  \label{HPT_M}
\end{equation}
where

\begin{equation}
\left( H_{\delta }\right) _{ij}=\delta _{ij}(V_{j}-V_{j}^{\ast })/2=i\text{Im%
}\left( V_{j}\right) \delta _{ij}  \label{H_delta}
\end{equation}%
and $H_{\gamma }$ ($H_{\alpha }$)\ is an $N_{1}$\ ($N_{2}$) dimension square
matrix. The matrices $H_{\gamma }$, $H_{\alpha }$, and $H_{\alpha \beta }$
are all real Hermitian while $H_{\gamma \alpha }$ is real. We can see that
Hamiltonian $H_{\mathcal{PT}}$\ describes an arbitrary real Hermitian graph
with parity-symmetry\ as defined in Eq. (\ref{P}) combining with the on-site
$\mathcal{PT}$-symmetric potentials $H_{\delta }$. Thus $H_{\mathcal{PT}}$
satisfies $\left[ \mathcal{PT}\text{, }H_{\mathcal{PT}}\right] =0$.

Introducing the linear transformation
\begin{subequations}
\label{j_AB}
\begin{eqnarray}
\left\vert j\right\rangle _{A} &=&\left\{
\begin{array}{cc}
\left\vert j\right\rangle _{c}, & \left( j\in \left[ 1,N_{1}\right] \right)
\\
(\left\vert j\right\rangle _{c}+\overline{\left\vert j\right\rangle }_{c})/%
\sqrt{2}, & \left( j-N_{1}\in \left[ 1,N_{2}\right] \right)%
\end{array}%
\right.  \label{j_AB_A} \\
\left\vert j\right\rangle _{B} &=&(\left\vert j\right\rangle _{c}-\overline{%
\left\vert j\right\rangle }_{c})/\sqrt{2},\text{ }\left( j-N_{1}\in \left[
1,N_{2}\right] \right)  \label{j_AB_B}
\end{eqnarray}%
one can rewrite the matrix of Eq. (\ref{HPT_M}) in the basis $\left\{
\left\vert j\text{, }j\in \left[ 1,N_{1}+N_{2}\right] \right\rangle
_{A}\right. $, $\left. \left\vert j\text{, }j\in \left[ N_{1}+1,N_{1}+N_{2}%
\right] \right\rangle _{B}\right\} $ as the form

\end{subequations}
\begin{equation}
H_{\mathcal{PT}}=\left(
\begin{array}{ccc}
H_{\gamma } & \sqrt{2}H_{\gamma \alpha } & 0 \\
\sqrt{2}H_{\gamma \alpha }^{\dagger } & H_{\alpha }+H_{\alpha \beta } &
H_{\delta } \\
0 & H_{\delta } & H_{\alpha }-H_{\alpha \beta }%
\end{array}%
\right)  \label{H_antidegger}
\end{equation}%
Obviously, it is the special case of Eq. (\ref{H_C_Matrix_form}), where%
\begin{eqnarray}
H_{A} &=&\left(
\begin{array}{cc}
H_{\gamma } & \sqrt{2}H_{\gamma \alpha } \\
\sqrt{2}H_{\gamma \alpha }^{\dagger } & H_{\alpha }+H_{\alpha \beta }%
\end{array}%
\right) , \\
H_{B} &=&H_{\alpha }-H_{\alpha \beta }, \\
H_{AB} &=&\left(
\begin{array}{c}
0 \\
H_{\delta }%
\end{array}%
\right) ,H_{BA}=\left(
\begin{array}{cc}
0 & H_{\delta }%
\end{array}%
\right) ,
\end{eqnarray}%
or equivalently in the explicit form as
\begin{eqnarray}
\left( H_{A}\right) _{mn}\left. =\right. _{A}\left\langle m\right\vert H_{%
\mathcal{PT}}\left\vert n\right\rangle _{A} &=&\left( H_{A}\right)
_{nm}^{\ast }, \\
\left( H_{B}\right) _{mn}\left. =\right. _{B}\left\langle m\right\vert H_{%
\mathcal{PT}}\left\vert n\right\rangle _{B} &=&\left( H_{B}\right)
_{nm}^{\ast }, \\
\left( H_{AB}\right) _{mn}\left. =\right. _{A}\left\langle m\right\vert H_{%
\mathcal{PT}}\left\vert n\right\rangle _{B} &=&i\text{Im}\left( V_{m}\right)
\delta _{m-N_{1},n}, \\
\left( H_{BA}\right) _{mn}\left. =\right. _{B}\left\langle m\right\vert H_{%
\mathcal{PT}}\left\vert n\right\rangle _{A} &=&i\text{Im}\left( V_{n}\right)
\delta _{m,n-N_{1}}.
\end{eqnarray}%
Therefore, the cluster $H_{\mathcal{PT}}$\ acts as a Hermitian scattering
center. This result is independent of the magnitudes of the hopping
integrals and the potentials, also the reality of the spectrum of $H_{%
\mathcal{PT}}$.

\section{Illustration}

\label{sec_illus}


\begin{figure}[tbp]
\includegraphics[ bb=66 290 550 745, width=8.5 cm, clip]{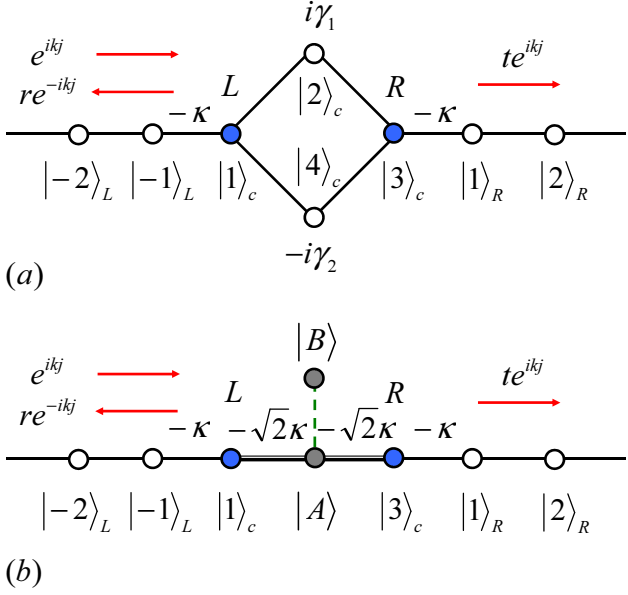}
\caption{(Color online) Schematic illustration for the exemplified
system.\ (a) $4$-site non-Hermitian scattering center configuration, which
consists of two on-site imaginary potentials $i\gamma _{1}$\ and $-i\gamma
_{2}$\ connecting to two semi-infinite chains as the waveguides at the joint
sites $\left\vert 1\right\rangle _{c}$\ and $\left\vert 3\right\rangle _{c}$, with the hopping strength $-\kappa $. (b) The equivalent Hamiltonian of $H_{C}$\ [Eq. (\ref{H_eg})], which is obtained under the linear
transformation of Eq. (\ref{A_and_B}). The hopping strengths between the site
$\left\vert A\right\rangle $\ and $\left\vert 1\right\rangle _{c}$, $\left\vert 3\right\rangle _{c}$\ are both $-\sqrt{2}\kappa $. The dashed
(green) line represents the effective hopping between sites $\left\vert
A\right\rangle $\ and $\left\vert B\right\rangle \ $which is pure imaginary $i\left( \gamma _{1}+\gamma _{2}\right) /2$, with both potentials on $\left\vert A\right\rangle $\ and $\left\vert B\right\rangle $ being $i\left(
\gamma _{1}-\gamma _{2}\right) /2$. In the case of $\gamma
_{1}=\gamma _{2}$, it is shown that the non-Hermitian scattering center acts
as a Hermitian one, preserving the Dirac probability current.} \label{fig2}
\end{figure}


We consider a simple $4$-site non-Hermitian scattering center which is
illustrated schematically in Fig. \ref{fig2}(a). The Hamiltonian of the
whole system of Eq. (\ref{H}) can be written as
\begin{eqnarray}
H_{L} &=&-\kappa \sum_{j=-1}^{-\infty }\left\vert j\right\rangle
_{L}\left\langle j-1\right\vert -\kappa \left\vert -1\right\rangle
_{Lc}\left\langle 1\right\vert +\text{H.c.}, \\
H_{R} &=&-\kappa \sum_{j=1}^{+\infty }\left\vert j\right\rangle
_{R}\left\langle j+1\right\vert -\kappa \left\vert 1\right\rangle
_{Rc}\left\langle 3\right\vert +\text{H.c.},
\end{eqnarray}%
where the joints of the scattering center are $L=1$, $R=3$ and%
\begin{eqnarray}
H_{C} &=&-\kappa \left( \left\vert 1\right\rangle _{c}\left\langle
2\right\vert +\left\vert 2\right\rangle _{c}\left\langle 3\right\vert
+\left\vert 3\right\rangle _{c}\left\langle 4\right\vert +\left\vert
4\right\rangle _{c}\left\langle 1\right\vert +\text{H.c.}\right)   \notag \\
&&+i\gamma _{1}\left\vert 2\right\rangle _{c}\left\langle 2\right\vert
-i\gamma _{2}\left\vert 4\right\rangle _{c}\left\langle 4\right\vert .
\end{eqnarray}%
Note that here we consider a non-$\mathcal{PT}$-symmetric model without
losing the generality. The Bethe ansatz wavefunction has the form%
\begin{equation}
\left\vert \phi _{k}\right\rangle =\sum_{j=-1}^{-\infty }f_{j}\left\vert
j\right\rangle _{L}+\sum_{j=1}^{4}h_{j}\left\vert j\right\rangle
_{c}+\sum_{j=1}^{+\infty }f_{j}\left\vert j\right\rangle _{R},
\end{equation}%
where $f_{j}$\ is in form of Eq. (\ref{f_i}). Taking $\kappa =1$,\ the
explicit form of Schr\"{o}dinger equations are

\begin{eqnarray}
-f_{-1}-h_{2}-h_{4} &=&Eh_{1},  \notag \\
-f_{1}-h_{2}-h_{4} &=&Eh_{3},  \notag \\
-h_{1}-h_{3} &=&\left( E-i\gamma _{1}\right) h_{2}, \\
-h_{1}-h_{3} &=&\left( E+i\gamma _{2}\right) h_{4},  \notag \\
E &=&-2\cos k.  \notag
\end{eqnarray}%
the continuity of the wavefunctions demands

\begin{equation}
h_{1}=1+r,h_{3}=t,  \label{h1h3}
\end{equation}%
The corresponding transmission and reflection coefficients have the form

\begin{eqnarray}
r &=&\frac{\zeta \cos k-1}{e^{-ik}-\zeta }e^{ik}, \\
t &=&-\frac{i\zeta \sin k}{e^{-ik}-\zeta }e^{ik},
\end{eqnarray}%
where%
\begin{equation}
\zeta =\frac{1}{\cos k+i\gamma _{1}/2}+\frac{1}{\cos k-i\gamma _{2}/2}.
\end{equation}%
Straightforward algebra shows that

\begin{equation}
\left\vert r\right\vert ^{2}+\left\vert t\right\vert ^{2}=1-\frac{2\text{Im}%
\left( \zeta \right) \sin k}{1+\left\vert \zeta \right\vert ^{2}-2\text{Re}%
\left( \zeta \right) \cos k+2\text{Im}\left( \zeta \right) \sin k},
\end{equation}%
which indicates the current is conserved when $\zeta $ is real, i.e., $%
\gamma _{1}=\gamma _{2}$.

Alternatively, taking the linear transformation
\begin{subequations}
\label{A_and_B}
\begin{eqnarray}
\left\vert A\right\rangle &=&\left( \left\vert 2\right\rangle
_{c}+\left\vert 4\right\rangle _{c}\right) /\sqrt{2},  \label{A_and_B_A} \\
\left\vert B\right\rangle &=&\left( \left\vert 2\right\rangle
_{c}-\left\vert 4\right\rangle _{c}\right) /\sqrt{2},  \label{A_and_B_B}
\end{eqnarray}%
the Hamiltonian $H_{C}$\ can be rewritten as

\end{subequations}
\begin{eqnarray}
H_{C} &=&-\sqrt{2}\kappa \left( \left\vert A\right\rangle _{c}\left\langle
1\right\vert +\left\vert A\right\rangle _{c}\left\langle 3\right\vert +\text{%
H.c.}\right)  \label{H_eg} \\
&&+i\frac{\left( \gamma _{1}-\gamma _{2}\right) }{2}\left( \left\vert
A\right\rangle \left\langle A\right\vert +\left\vert B\right\rangle
\left\langle B\right\vert \right)  \notag \\
&&+i\frac{\left( \gamma _{1}+\gamma _{2}\right) }{2}\left( \left\vert
A\right\rangle \left\langle B\right\vert +\left\vert B\right\rangle
\left\langle A\right\vert \right) ,  \notag
\end{eqnarray}%
as illustrated schematically in Fig. \ref{fig2}(b). It depicts a scattering
configuration with single side coupling site which has been systematically
studied in the Hermitian regime \cite{JLtrap}. Obviously, when $\gamma
_{1}=\gamma _{2}=\gamma $, it is a simple example of a real Hermitian graph
with parity-symmetry combining with the on-site $\mathcal{PT}$-symmetric
potentials and admits the current preserving.

Accordingly, the transmission probability (coefficient) has the form%
\begin{equation}
T\left( k\right) =\frac{\sin ^{2}\left( 2k\right) }{\sin ^{2}\left(
2k\right) +\left( \cos ^{2}\left( k\right) -\gamma ^{2}/4\right) ^{2}},
\label{T_k}
\end{equation}%
which has peculiar feature in contrast to that of Hermitian scattering
center. As a comparison, we write the transmission probability for the real
side coupling by substituting $\gamma $\ with $i\gamma $, i.e.,%
\begin{equation}
T^{\prime }\left( k\right) =\frac{\sin ^{2}\left( 2k\right) }{\sin
^{2}\left( 2k\right) +\left( \cos ^{2}\left( k\right) +\gamma ^{2}/4\right)
^{2}}.  \label{T'_k}
\end{equation}%
It can be observed that, (i) both of them have the common total reflection
points, $T\left( \pi /2\right) =T^{\prime }\left( \pi /2\right) =0$; (ii) $%
T\left( k\right) =1$ at the resonance condition $\gamma =2\left\vert \cos
k\right\vert $, while $T^{\prime }\left( k\right) $ is always less than $1$
within the\ whole range of $\gamma $.

\section{Summary and discussion}

\label{sec_summary} We have proposed an anti-Hermitian coupled two Hermitian
graphs as the scattering center, which has been shown to act as a Hermitian
graph, preserving the traditional probability. This conclusion can be
applied to the non-Hermitian scattering center which consists of pairs of $%
\mathcal{PT}$-symmetric on-site potentials. This fact indicates the balanced
gain and loss can result in the Hermiticity of the scattering center. Our
results can give a good prediction for the transmission and reflection
coefficients of linear waves scattered at the $\mathcal{PT}$-symmetric
defects in the experiment. The recent observation of breaking of $\mathcal{PT%
}$ symmetry in coupled optical waveguides \cite{AGuo,TK,CERuter} may pave
the way to demonstrate the result presented in this paper.

Finally, we would like to point that our conclusion can also apply to other
type of non-Hermitian scattering center. For instance, we can select $%
H_{\gamma }$, $H_{\alpha }$, $H_{\gamma \alpha }$ and $H_{\alpha \beta }$
being all Hermitian instead of real Hermitian in Eq. (\ref{HPT_M}), and we
note that $H_{\mathcal{PT}}$ is no longer $\mathcal{PT}$-symmetric if the
hoppings are not all real, with $H_{\gamma }$, $H_{\alpha }$, $H_{\gamma
\alpha }$ and $H_{\alpha \beta }$ being Hermitian, we could also select $H_{%
\mathcal{PT}}$ as

\begin{equation}
H_{\mathcal{PT}}=\left(
\begin{array}{ccc}
H_{\gamma } & H_{\gamma \alpha } & H_{\gamma \alpha } \\
H_{\gamma \alpha }^{\dagger } & H_{\alpha }+H_{\delta } & H_{\alpha \beta }
\\
H_{\gamma \alpha }^{\dagger } & H_{\alpha \beta }^{\ast } & H_{\alpha
}-H_{\delta }%
\end{array}%
\right) .
\end{equation}%
which is also in form of Eq. (\ref{H_C_Matrix_form}) after the
transformation of Eq. (\ref{j_AB}) and exhibits the Hermitian behavior.

\section{Appendix}

In this Appendix, we will prove the relation of Eq. (\ref{Delta_ij}). For an
incident plane wave with real energy $E$, we obtain from Eq. (\ref{Delta})
and $H_{A}=H_{A}^{\dagger }$, $H_{B}=H_{B}^{\dagger }$ that%
\begin{equation}
\Delta ^{\dagger }=\left(
\begin{array}{cc}
H_{A}-E & -H_{AB} \\
H_{AB}^{\dag } & H_{B}-E%
\end{array}%
\right)  \label{Delta_dagger}
\end{equation}%
Considering the block matrix $\Delta $ and $\Delta ^{\dagger }$, when $%
\left( H_{B}-E\right) $ is invertible, employing the Leibniz formula, we have%
\begin{eqnarray}
\det \left( \Delta \right) &=&\det \left( H_{B}-E\right) \det [\left(
H_{A}-E\right)  \label{appdendix_DET} \\
&&-H_{AB}\left( H_{B}-E\right) ^{-1}\left( -H_{AB}\right) ^{\dag }],  \notag
\end{eqnarray}%
and also%
\begin{eqnarray}
\det \left( \Delta ^{\dagger }\right) &=&\det \left( H_{B}-E\right) \det
[\left( H_{A}-E\right)  \label{appdendix_DET_dagger} \\
&&-\left( -H_{AB}\right) \left( H_{B}-E\right) ^{-1}H_{AB}^{\dag }].  \notag
\end{eqnarray}%
Then we have

\begin{equation}
\det \left( \Delta \right) =\det \left( \Delta ^{\dagger }\right) =\left[
\det \left( \Delta ^{T}\right) \right] ^{\ast }=\left[ \det \left( \Delta
\right) \right] ^{\ast },
\end{equation}%
i.e., $\det \left( \Delta \right) $ is real. Such feature arises from the
special structure of the matrix $\Delta $ in the form

\begin{equation}
\left(
\begin{array}{cc}
\mathcal{A} & \mathcal{C} \\
-\mathcal{C}^{\dag } & \mathcal{B}%
\end{array}%
\right) ,  \label{sample}
\end{equation}%
with $\mathcal{A}$ and $\mathcal{B}$ being Hermitian matrices.

Matrix $M_{ij}$\ ($i,j\in \left[ 1,N_{A}\right] $) is obtained from $\Delta $%
\ by eliminating its $i$th row and $j$th column, which has the form

\begin{equation}
\left(
\begin{array}{cc}
\mathcal{A}^{\prime } & \mathcal{C}^{\prime } \\
\mathcal{D}^{\prime } & \mathcal{B}%
\end{array}%
\right) ,
\end{equation}%
and accordingly $\left( M_{ji}\right) ^{\dag }$\ has the form

\begin{equation}
\left(
\begin{array}{cc}
\mathcal{A}^{\prime } & -\mathcal{C}^{\prime } \\
-\mathcal{D}^{\prime } & \mathcal{B}%
\end{array}%
\right) .
\end{equation}%
Here, $\mathcal{A}^{\prime }$\ is the matrix by eliminating the $i$th row
and $j$th column from $\mathcal{A}$, while $\mathcal{C}^{\prime }$ ($%
\mathcal{D}^{\prime }$) is the matrix by eliminating the $i$th row ($j$th
column) from $\mathcal{C}$ ($\mathcal{D}$). By similar procedure in
obtaining Eqs. (\ref{appdendix_DET}) and (\ref{appdendix_DET_dagger}), we
have%
\begin{equation}
\det \left( M_{ij}\right) =\det [\left( M_{ji}\right) ^{\dag }],
\end{equation}%
and then
\begin{equation}
\det \left( M_{ij}\right) =\left[ \det \left( M_{ji}\right) \right] ^{\ast }.
\end{equation}%
Together with Eq. (\ref{minor}), we have $\left( \Delta ^{-1}\right)
_{ij}=\left( \Delta ^{-1}\right) _{ji}^{\ast }$, and yield Eq. (\ref%
{inverse_delta_rela}).

\acknowledgments We acknowledge the support of the CNSF (Grant No. 10874091)
and National Basic Research Program (973 Program) of China under Grant No.
2012CB921900.


\begin{thebibliography}{99}
\bibitem{Bender 98} C. M. Bender, and S. Boettcher, Phys. Rev. Lett. \textbf{%
80}, 5243 (1998).

\bibitem{Bender 99} C. M. Bender, S. Boettcher, and P. N. Meisinger, J.
Math. Phys. \textbf{40}, 2201 (1999).

\bibitem{Dorey 01} P. Dorey, C. Dunning, and R. Tateo, J. Phys. A: Math.
Gen. \textbf{34}, L391 (2001); P. Dorey, C. Dunning, and R. Tateo, J. Phys.
A: Math. Gen. \textbf{34}, 5679 (2001).

\bibitem{Bender 02} C. M. Bender, D. C. Brody, and H. F. Jones, Phys. Rev.
Lett. \textbf{89}, 270401 (2002).

\bibitem{A.M43} A. Mostafazadeh, J. Math. Phys. \textbf{43}, 3944 (2002).

\bibitem{A.M} A. Mostafazadeh and A. Batal, J. Phys. A: Math. Gen. \textbf{37%
}, 11645 (2004).

\bibitem{A.M36} A. Mostafazadeh, J. Phys. A: Math. Gen. \textbf{36}, 7081
(2003).

\bibitem{Jones} H. F. Jones, J. Phys. A: Math. Gen. \textbf{38}, 1741 (2005).

\bibitem{A.M38} A. Mostafazadeh, J. Phys. A: Math. Gen. \textbf{38}, 6557
(2005).

\bibitem{A.M391} A. Mostafazadeh, J. Phys. A: Math. Gen. \textbf{39}, 10171
(2006).

\bibitem{A.M392} A. Mostafazadeh, J. Phys. A: Math. Gen. \textbf{39}, 13495
(2006).

\bibitem{Joglekar82} Y. N. Joglekar, D. Scott, M. Babbey, and A. Saxena,
Phys. Rev. A \textbf{82}, 030103(R) (2010).

\bibitem{Joglekar83} D. D. Scott and Y. N. Joglekar, Phys. Rev. A \textbf{83}%
, 050102(R) (2011).

\bibitem{JLPT} L. Jin and Z. Song, Phys. Rev. A \textbf{80}, 052107 (2009).

\bibitem{Bendix} O. Bendix, R. Fleischmann, T. Kottos and B. Shapiro, Phys.
Rev. Lett. \textbf{103}, 030402 (2009).

\bibitem{Keya} K. Zhou, Z. Guo, J. Wang and S. Liu Opt. Lett. \textbf{35},
2928 (2010).

\bibitem{YDChong} Y. D. Chong, Li Ge, Hui Cao and A. D. Stone, Phys. Rev.
Lett. \textbf{105}, 053901 (2010).

\bibitem{LonghiLaser} S. Longhi, Phys. Rev. A \textbf{82}, 031801(R) (2010);
Phys. Rev. Lett. \textbf{105}, 013903 (2010).

\bibitem{JLCTP} L. Jin and Z. Song, Phys. Rev. A \textbf{81}, 032109 (2010);
Commun. Theor. Phys. \textbf{54}, 73 (2010).

\bibitem{ZAhmed} Z. Ahmed, Phys. Lett. A \textbf{324,} 152 (2004).

\bibitem{MZnojil} M. Znojil, J. Phys. A: Math. Gen. \textbf{39,} 13325
(2006); Phys. Rev. D \textbf{78}, 025026 (2008); Phys. Rev. D \textbf{80},
045009 (2009).

\bibitem{HJones} H. F. Jones, Phys. Rev. D \textbf{76}, 125003 (2007); Phys.
Rev. D \textbf{78}, 065032 (2008).

\bibitem{Kivshar} S. V. Dmitriev, S. V. Suchkov, A. A. Sukhorukov, and Y. S.
Kivshar Phys. Rev. A \textbf{84}, 013833 (2011).

\bibitem{JLtrap} L. Jin and Z. Song, Phys. Rev. A \textbf{81}, 022107 (2010).

\bibitem{AGuo} A. Guo, G. J. Salamo, D. Duchesne, R.Morandotti, M.
Volatier-Ravat, V. Aimez, G. A. Siviloglou, and D. N. Christodoulides, Phys.
Rev. Lett. \textbf{103}, 093902 (2009).

\bibitem{TK} T. Kottos, Nature Phys. \textbf{6}, 166 (2010).

\bibitem{CERuter} C. E. R\"{u}ter, K. G. Makris, R. El-Ganainy, D. N.
Christodoulides, M. Segev, and D. Kip, Nature Phys. \textbf{6}, 192 (2010).
\end{thebibliography}
\end{document}